\documentclass[
 reprint,
 amsmath,
 amssymb,
 aps,
 prl,
 twocolumn,
 nolongbibliography
]{revtex4-2}

\usepackage{multirow}
\usepackage{amsmath,amsfonts,amssymb}
\usepackage{dcolumn}

\usepackage{graphicx}
\usepackage{bm}
\usepackage{gensymb}
\usepackage[utf8]{inputenc}
\usepackage{threeparttable, tablefootnote}
\usepackage{booktabs}
\usepackage{csquotes}
\usepackage[export]{adjustbox}
\usepackage{placeins}
\usepackage{setspace}
\usepackage{caption}
\usepackage{subcaption}
\usepackage[usenames,dvipsnames]{color}
\usepackage{cancel}
\usepackage{nicefrac}

\DeclareCaptionJustification{myjust}{\fulljustify}
\makeatletter
\providecommand\fulljustify{%
  \let\\\@centercr
  \leftskip\z@%
  \rightskip\z@%
  \parfillskip\z@\@plus 1fill\relax%
}

\usepackage{array,mathtools,booktabs}
\newcolumntype{C}{>{$}c<{$}}
\AtBeginDocument{
\heavyrulewidth=.08em
\lightrulewidth=.05em
\cmidrulewidth=.03em
\belowrulesep=.65ex
\belowbottomsep=0pt
\aboverulesep=.4ex
\abovetopsep=0pt
\cmidrulesep=\doublerulesep
\cmidrulekern=.5em
\defaultaddspace=.5em
}

\usepackage{hyperref}
\usepackage{siunitx}
\usepackage{todonotes}
\presetkeys{todonotes}{inline,backgroundcolor=yellow}{}

\DeclareSIUnit\year{yr}
\DeclareSIUnit\sinv{s^{-1}}
\DeclareSIUnit\torr{Torr}
\DeclareSIUnit\sinvovtorr{s^{-1} \cdot Torr^{-1}}
\DeclareSIUnit\literpers{L/s}
\DeclareSIUnit\uCi{\text{$\mu$}Ci}

\newcommand{\reportednumber}[1]{\textcolor{Black}{#1}} 
\newcommand{\systemnumber}[1]{\textcolor{Black}{#1}} 

\newcommand{\reportingdfiveovtwolifetime}[0]{$303.8(1.5)$ ms}
\newcommand{\reportingdfiveovtwodecayrate}[0]{\SI{3.291\pm0.017}{\sinv}}
\newcommand{\reportingdthreeovtwolifetime}[0]{\SI{642\pm9}{\ms}}
\newcommand{\reportingdthreeovtwodecayrate}[0]{\SI{1.56\pm0.02}{\sinv}}

\newcommand{\tref}[1]{Table~\ref{#1}}

\newcommand{\preivousexpdfivelifetimelowerbond}[0]{\SI{232\pm4}{\ms}}

\newcommand{\separationrf}[0]{\SI{6} {\mm}}
\newcommand{\separationendcap}[0]{\SI{15} {\mm}}

\newcommand{\Ratwotwentysixhalflife}[0]{\SI{1600}{\year}}
\newcommand{\Ratwotwentyfourhalflife}[0]{\SI{3.6}{\day}}
\newcommand{\Rntwotwentytwohalflife}[0]{\SI{3.8}{\day}}
\newcommand{\Rntwotwentyhalflife}[0]{\SI{55.6}{\s}}
\newcommand{\previouschamberpressure}[0]{\SI{3e-10}{\torr}}
\newcommand{\Rasourceamount}[0]{\SI{10}{\uCi}}
\newcommand{\Thsourceamount}[0]{\SI{25}{\uCi}}

\newcommand{\tenmsbrightcts}[0]{126}
\newcommand{\tenmsdarkcts}[0]{9.5}
\newcommand{\tenmsthreshold}[0]{35}
\newcommand{\statedetecttime}[0]{\SI{10} {\ms}}
\newcommand{\reportingdatapressure}[0]{\SI{5e-11} {\torr}}

\newcommand{\opticalpumptime}[0]{\SI{0.5} {\ms}}

\newcommand{\branchratiodfiveovtwotosoneovtwo}[0]{\SI{10.930\pm0.013}{}}

\newcommand{\totaldfiveshiftdecayrate}[0]{7}
\newcommand{\totaldfiveshiftuncdecayrate}[0]{17}

\newcommand{\totaldthreeshiftdecayrate}[0]{10}
\newcommand{\totaldthreeshiftuncdecayrate}[0]{20}

\newcommand{\dfivefitreducedchisquare}[0]{1.56}
\newcommand{\dthreefitreducedchisquare}[0]{1.11}

\newcommand{\statisticaldfiveovtwodecayrate}[0]{\SI{3.284\pm0.014}{\sinv}}
\newcommand{\statisticaldthreeovtwodecayrate}[0]{\SI{1.54\pm0.02}{\sinv}}
\newcommand{\statisticaldfiveovtwodecayrateunc}[0]{14}
\newcommand{\statisticaldthreeovtwodecayrateunc}[0]{20}

\newcommand{\elasticcollisionrate}[0]{\SI{1.1\pm0.3 e-4}{\sinv}}

\newcommand{\dfiveelasticcollisionshifttotaldecayrate}[0]{$7.3(1.9) \times 10^{-3}~\text{s}^{-1}$}
\newcommand{\dfiveelasticcollisionshiftdecayrate}[0]{7.3}
\newcommand{\dfiveelasticcollisionshiftdecayrateunc}[0]{1.9}

\newcommand{\dthreeelasticcollisionshifttotaldecayrate}[0]{\SI{1.2\pm0.3 e-2}{\sinv}}
\newcommand{\dthreeelasticcollisionshiftdecayrate}[0]{12}
\newcommand{\dthreeelasticcollisionshiftdecayrateunc}[0]{3}

\newcommand{\dfiveinelasticcollisionpressureone}[0]{\SI{1.1e-10}{\torr}}
\newcommand{\dfiveinelasticcollisionpressuretwo}[0]{\SI{2.2e-10}{\torr}}
\newcommand{\measureddfiveinelasticcollisionratevspressure}[0]{\SI{1\pm17 e7}{\sinvovtorr}}

\newcommand{\dfiveinelasticcollisionshifttotaldecayrate}[0]{$6(85) \times 10^{-4}~\text{s}^{-1}$}
\newcommand{\dfiveinelasticcollisionshiftdecayrate}[0]{-0.6}
\newcommand{\dfiveinelasticcollisionshiftdecayrateunc}[0]{8.5}
\newcommand{\dfiveinelasticcollisionshiftdecayrateuncunit}[0]{\SI{8.5 e-3}{\sinv}}

\newcommand{\threetofivemixingrate}[0]{\SI{1\pm2 e-4}{\sinv}}
\newcommand{\threetofivemixingshiftindecayrate}[0]{\SI{3\pm6 e-4}{\sinv}}

\newcommand{\dthreeinelasticcollisionshiftdecayrate}[0]{0.3}
\newcommand{\dthreeinelasticcollisionshiftdecayrateunc}[0]{8.6}

\newcommand{\dfiveinputdecayrateanalysisuncertaintynounit}[0]{0.09}
\newcommand{\dthreeinputdecayrateanalysisuncertaintynounit}[0]{0.06}

\newcommand{\blackbodydfivedecayrateshiftnounit}[0]{-0.017}
\newcommand{\blackbodydfivedecayrateshiftuncnounit}[0]{0.005}
\newcommand{\blackbodydthreedecayrateshiftnounit}[0]{0.05}
\newcommand{\blackbodydthreedecayrateshiftuncnounit}[0]{0.01}

\begin{document}

\newcommand{\UCSBAffiliation}[0]{Department of Physics, University of California, Santa Barbara, California 93106, USA}

\title{Lifetimes of the Metastable $6\mathrm{d}\, ^{2}\mathrm{D}_{5/2}$ and $6\mathrm{d}\, ^{2}\mathrm{D}_{3/2}$ States of Ra$^+$}

\author{Haoran Li}
\email{hli836@ucsb.edu}
\affiliation{\UCSBAffiliation}
\author{Huaxu Dan}
\affiliation{\UCSBAffiliation}
\author{Mingyu Fan}
\affiliation{\UCSBAffiliation}
\author{Spencer Kofford}
\affiliation{\UCSBAffiliation}
\author{Robert Kwapisz}
\affiliation{\UCSBAffiliation}
\author{Roy A. Ready}
\affiliation{\UCSBAffiliation}
\author{Akshay Sawhney}
\affiliation{\UCSBAffiliation}
\author{Merrell Brzeczek}
\affiliation{\UCSBAffiliation}
\author{Craig Holliman}
\affiliation{\UCSBAffiliation}
\author{Andrew M. Jayich}
\affiliation{\UCSBAffiliation}

\author{S. G. Porsev}
\affiliation{Department of Physics and Astronomy, University of Delaware, Newark, Delaware 19716, USA}
\author{M. S. Safronova}
\email{msafrono@physics.udel.edu}
\affiliation{Department of Physics and Astronomy, University of Delaware, Newark, Delaware 19716, USA}
\affiliation{Joint Quantum Institute, National Institute of Standards and Technology and the University of Maryland, College Park, Maryland 20742, USA}

\date{\today}

\begin{abstract}

We report lifetime measurements of the metastable $6\mathrm{d}\, ^{2}\mathrm{D}_{5/2}$ and $6\mathrm{d}\, ^{2}\mathrm{D}_{3/2}$ states of Ra$^+$. 
The measured lifetimes, $\tau_{5} = $ \reportednumber{\reportingdfiveovtwolifetime}
and $\tau_{3} = $ \reportednumber{\reportingdthreeovtwolifetime},
are important for optical frequency standards and for benchmarking high-precision relativistic atomic theory.  Independent of the reported measurements,  the D state lifetimes were calculated using the coupled-cluster single double triple method, in which the coupled-cluster equations for both core and valence triple excitations were solved iteratively.  The method was designed for precise prediction of atomic properties, especially for heavy elements, where relativistic and correlation corrections become large, making their treatment more challenging. 
This Letter presents the first tests of the method for transition properties. Our prediction agrees with experimental values within the uncertainties. 
The ability to accurately predict the atomic properties of heavy elements is important for many applications, from tests of fundamental symmetries to the development of optical clocks.
\end{abstract}

\maketitle

The radioactive elements at the bottom of the periodic table are intriguing both for science and technology \cite{ArrowsmithKron2024}.  However, many isotopes are challenging for experimentation due to their short half-lives.  Therefore, accurate theoretical predictions of transition frequencies and electronic state lifetimes can provide helpful guidance for experiments.  But unfortunately the large atomic numbers of heavy elements make accurate calculations notoriously difficult because of both electron correlations and relativistic effects.  To accurately calculate their properties the coupled-cluster single double triple method (CCSDT) was developed \cite{CCSDT}.  It was applied to extract nuclear moments from the hyperfine structure of $^{229}$Th in \cite{CCSDT}. Here, we present the first test of this approach comparing the \textit{ab initio} theoretical values with experimental results of the metastable D state lifetimes of Ra$^+$. 

\begin{figure}
    \centering
    \includegraphics[width=8.6cm]{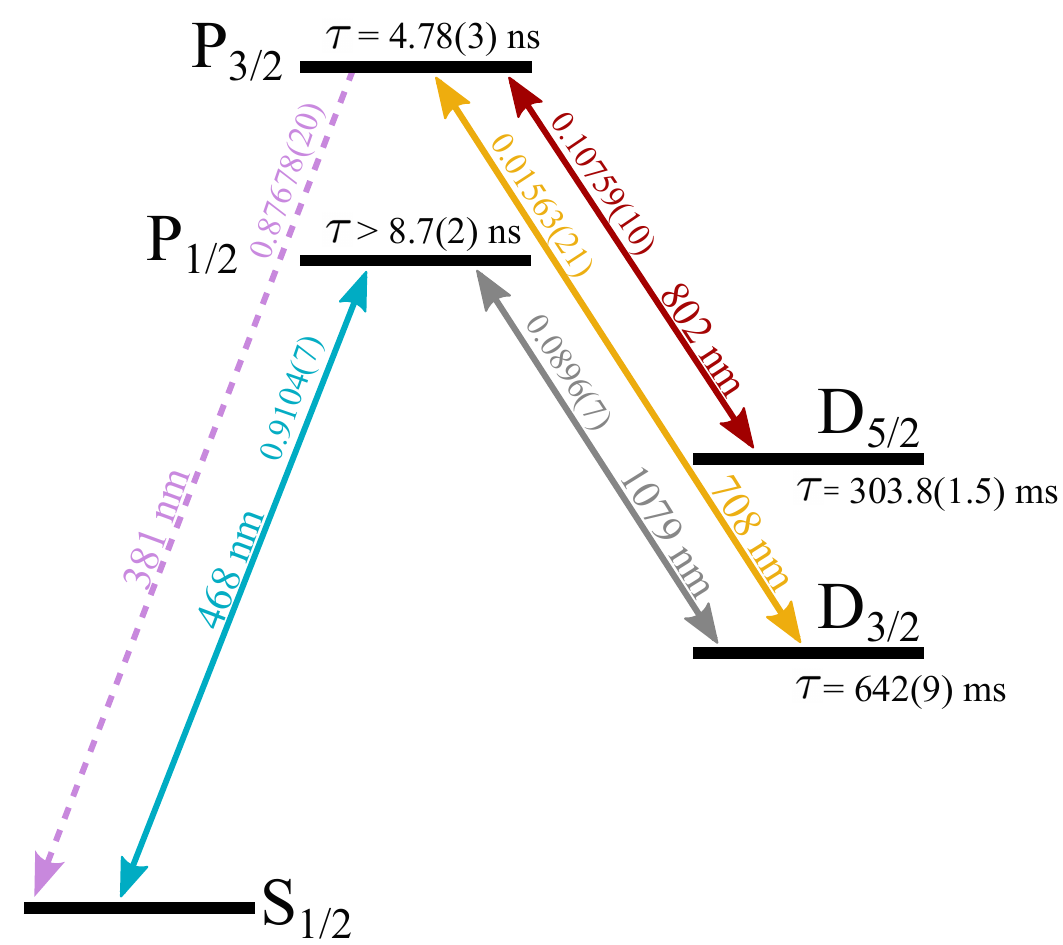}
    \captionsetup{justification=myjust, singlelinecheck=false}
    \caption{Low-lying Ra$^+$ level structure, with experimentally measured branching ratios and lifetimes~\cite{Fan2019, Fan2019a, Fan2022, Kofford2025}. The solid lines indicate the transitions driven in this measurement.}
    \label{fig:ra_ion_level}
\end{figure}

We report a precision measurement of the $6\mathrm{d}\, ^{2}\mathrm{D}_{5/2}$ state lifetime of Ra$^+$, $\tau_{5} =$ \reportednumber{\reportingdfiveovtwolifetime}, improving on a previous lower bound, $\tau_{5} \geq \reportednumber{\preivousexpdfivelifetimelowerbond}$~\cite{Versolato2011a}, and the first measurement of the  $6\mathrm{d}\, ^{2}\mathrm{D}_{3/2}$ state lifetime,  $\tau_{3} = \reportednumber{\reportingdthreeovtwolifetime}$, see Fig.~\ref{fig:ra_ion_level}.   The measurement precision is sufficient to support a test between experiment and the CCSDT theory predictions, which were estimated to be accurate to 1\%. 

The metastable D states of Ra$^+$ both support optical clock transitions.  The radium ion is appealing for optical clocks both for achieving very high precision and for realizing transportable devices \cite{Holliman2022}.  Its high mass and low charge to mass ratio reduce leading systematic uncertainties arising from the second-order Doppler effect \cite{Brewer2019}.  With $^{225}$Ra$^+$ (nuclear spin $I=1/2$) for both the $\mathrm{S}_{1/2} \leftrightarrow \mathrm{D}_{5/2}$ and $\mathrm{S}_{1/2} \leftrightarrow \mathrm{D}_{3/2}$, there are transitions which are first-order insensitive to magnetic field noise and both support optical clocks.  Because of the large hyperfine structure of $^{225}$Ra$^+$ it is possible to operate an optical clock with only laser light at 828 and 1079 nm using the $\mathrm{S}_{1/2} \leftrightarrow \mathrm{D}_{3/2}$ transition for the clock  \cite{Holliman2023}.  Requiring only low-power IR light could facilitate the use of integrated photonics that would help enable a transportable clock.  Knowledge of both lifetimes is important for calculating clock performance.

In the theoretical calculations, we took into account triple excitations and nonlinear (NL) terms, evaluated the contribution of higher partial waves, and computed smaller effects such as the Breit interaction and quantum electrodynamical corrections. Both core and valence triple excitations were included on the same footing as single and double excitations, i.e., iteratively solving the equations for triple cluster amplitudes. The NL terms were included in the equations for single and double excitations. This approach represents the most accurate treatment of electronic correlations in heavy systems. 

This new method allowed us to reduce the uncertainty of the matrix elements $\langle 7\mathrm{s}\,^2\mathrm{S}_{1/2} ||Q|| 6\mathrm{d}\,^2\mathrm{D}_{3/2,5/2} \rangle$ to the level of 0.5\%. Performing several calculations with increasing complexity enables us to put an uncertainty bound on our values. Comparing the theoretical results with the precision measurements carried out in this Letter, we observe an excellent agreement, confirming not only the validity of our approach but also our estimate of theory uncertainty, which is crucial for the other cases where experiments are not yet available. This is a good test of our CCSDT approach for transition matrix elements and lifetimes, which are determined by the quality of the wave functions at a large distance from the nucleus.  

We note that accurate prediction of actinide properties is a very challenging task because of substantial core-valence correlations that need to be treated with a higher level of precision compared to lighter elements. 

This theoretical benchmark is important for other proposed clocks based on electric-quadrupole transitions in Cf$^{15+}$ and Cf$^{17+}$~\cite{Cf}. The uncertainty of predictions for clock transitions in these ions is largely dominated by the effect of triple excitations in the coupled-cluster part of the computation, which is tested in this Letter. 

{\it Experiment}---We laser cool single $^{224}\mathrm{Ra}^{+}$ (\reportednumber{\Ratwotwentyfourhalflife} half-life) ions in a linear Paul trap with rf electrodes separated by \systemnumber{\separationrf} and end cap electrodes separated by \systemnumber{\separationendcap}, described in Ref.~\cite{Fan2023a}. The trap is in a vacuum chamber with a background pressure of \systemnumber{\reportingdatapressure} measured with an ion gauge.

Lifetime measurements were initially attempted with $^{226}$Ra$^+$ (\reportednumber{\Ratwotwentysixhalflife} half-life) in the same ion trap with a background pressure of \systemnumber{\previouschamberpressure}. The pressure was likely limited by $^{222}$Rn (\reportednumber{\Rntwotwentytwohalflife} half-life) which was generated by the decay of the $^{226}$Ra source (\reportednumber{\Rasourceamount}).  The measured lifetimes were systematically shifted up, probably from collisions with $^{222}$Rn and trapped ions loaded from ionizing radiation.  We tested the strength of the background ionizing radiation by successfully loading Sr$^+$ from a Sr atomic beam without photoionization light.  These collisional effects were reduced by using $^{224}$Ra$^+$, which decays to $^{220}$Rn, which has a relatively short \reportednumber{\Rntwotwentyhalflife} half-life. The $^{224}$Ra$^+$ was loaded via photoionization from a $^{224}$Ra atomic beam generated from a $^{228}$Th source (\reportednumber{\Thsourceamount})~\cite{Fan2023a}. 

\begin{figure}
    \centering
    \includegraphics[width=8.6cm]{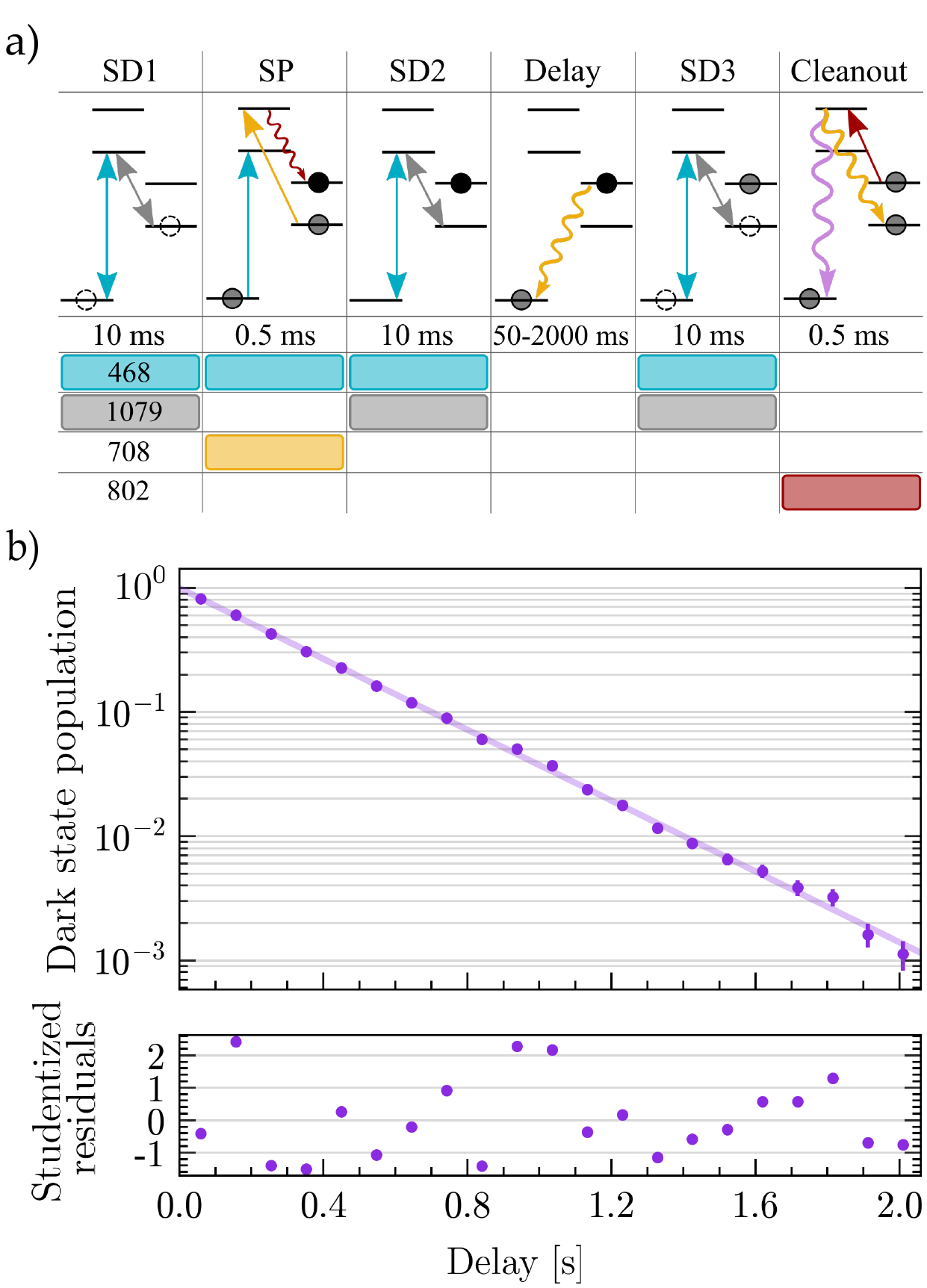}
    \captionsetup{justification=myjust, singlelinecheck=false}
    \caption{(a) $6\mathrm{d}\,^2\mathrm{D}_{5/2}$ state lifetime measurement pulse sequence, and (b) measured dark state population fit to exponential decay with $\chi^2_\nu = \reportednumber{\dfivefitreducedchisquare}$.}
    \label{fig:d5ov2_pulse_sequence_and_data}
\end{figure}

\begin{figure}
    \centering
    \includegraphics[width=8.6cm]{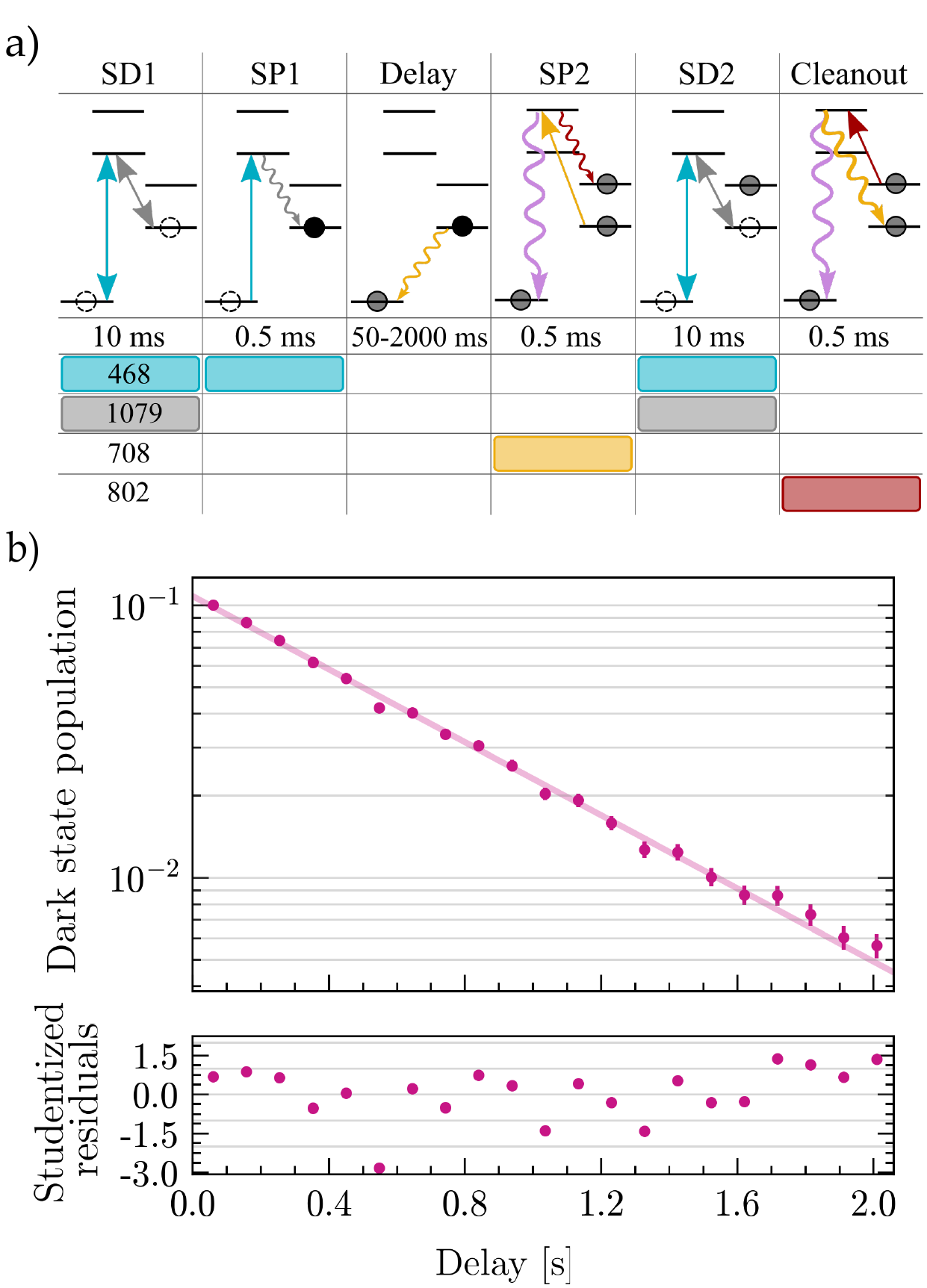}
    \captionsetup{justification=myjust, singlelinecheck=false}
    \caption{(a) $6\mathrm{d}\,^2\mathrm{D}_{3/2}$ state lifetime measurement pulse sequence, and (b) measured dark state population fit to exponential decay with $\chi^2_\nu = \reportednumber{\dthreefitreducedchisquare}$.}
    \label{fig:d3ov2_pulse_sequence_and_data}
\end{figure}

The measurement pulse sequences consist of optical pumping to the target D state, a variable delay time in the dark, and state detection; see Figs.~\ref{fig:d5ov2_pulse_sequence_and_data} (a) and~\ref{fig:d3ov2_pulse_sequence_and_data} (a). The measurements start with Doppler cooling and a state detection pulse (SD1) that confirms that the ion is cold and in a bright state. The ion scatters 468 nm photons when it is illuminated by the 468 and 1079 nm lasers and is in the $\mathrm{S}_{1/2}$ or the $\mathrm{D}_{3/2}$ ``bright'' states.  A fraction of the scattered photons are collected on a photomultiplier tube. If the ion is in a bright state, on average \systemnumber{\tenmsbrightcts} photons are detected during the \systemnumber{\statedetecttime}-long state detection. If the ion is in the $\mathrm{D}_{5/2}$ ``dark'' state, on average only \systemnumber{\tenmsdarkcts} photons are detected from background scattered light. Before the variable delay, we set a detection threshold of \systemnumber{\tenmsthreshold} photon counts to determine the ion's state. The dark state probability after the variable delay is determined using the maximum likelihood technique \cite{Fan2019a}. 

For the $\mathrm{D}_{5/2}$ lifetime measurement, the cleanout or state preparation (SP) might fail, in which case we reject measurements where SD1 is dark or SD2 is bright. We fit the data, see Fig. \ref{fig:d5ov2_pulse_sequence_and_data} (b), to an exponential decay, $p = ae^{-t/\tau_{5}}$, where $p$ is the $\mathrm{D}_{5/2}$ state population, $a$ is the amplitude, and $\tau_{5}$ is the lifetime of the $\mathrm{D}_{5/2}$ state. The fit gives $1/\tau_{5} = \reportednumber{\statisticaldfiveovtwodecayrate}$.

Because both the $\mathrm{D}_{3/2}$ and $\mathrm{S}_{1/2}$ states are bright states, for the $\mathrm{D}_{3/2}$ lifetime measurement we apply a \systemnumber{\opticalpumptime} 708 nm pulse (SP2) after the delay to optically pump \reportednumber{\branchratiodfiveovtwotosoneovtwo}\% of the $\mathrm{D}_{3/2}$ state population to the $\mathrm{D}_{5/2}$ state through the $\mathrm{P}_{3/2}$ state \cite{Fan2019a}. We reject measurements where SD1 is dark.  The data is fit to an exponential see Fig. \ref{fig:d3ov2_pulse_sequence_and_data} (b), which gives $1/\tau_{3} = \reportednumber{\statisticaldthreeovtwodecayrate}$.

Elastic and inelastic background gas collisions are the leading cause of systematic uncertainties for both lifetime measurements. Elastic collisions increase the kinetic energy of the ion, which can Doppler broaden transitions or remove the ion from the fiducial region, reducing the ion's photon scattering rate during state detection and shifting up the measured lifetimes. We measure elastic collision rates by preparing the ion in the $\mathrm{S}_{1/2}$ state and measuring the bright state probability versus delay time. The measured elastic collision rate, \reportednumber{\elasticcollisionrate} at \systemnumber{\reportingdatapressure}, shifts the $\mathrm{D}_{5/2}$ and $\mathrm{D}_{3/2}$ state decay rates down by \reportednumber{\dfiveelasticcollisionshifttotaldecayrate} and \reportednumber{\dthreeelasticcollisionshifttotaldecayrate}, respectively.

Fine structure mixing occurs when inelastic collisions transfer population between the $\mathrm{D}_{3/2}$ and $\mathrm{D}_{5/2}$ states. The transfer rate is $r_{53}$ from $\mathrm{D}_{5/2}$ to $\mathrm{D}_{3/2}$ and $r_{35}$ for the reverse process. For low collision rates we make the approximation $r_{35}\rightarrow0$ ($r_{53}\rightarrow0$) when the ion is initialized in the $\mathrm{D}_{5/2}$ ($\mathrm{D}_{3/2}$) state~\cite{supp}. Inelastic collisions can also quench the ion to its electronic ground state.  It is reasonable to assume that the quenching rate, $r_{\text{q}}$, is the same for both D states~\cite{Yu1997}.

For the $\mathrm{D}_{5/2}$ state, inelastic collisions shift the decay rate by $r_{53} + r_{\text{q}}$. To measure $r_{53} + r_{\text{q}}$, we made two $\mathrm{D}_{5/2}$ lifetime measurements at elevated pressures, \mbox{\systemnumber{\dfiveinelasticcollisionpressureone}} and \systemnumber{\dfiveinelasticcollisionpressuretwo}, and obtain $r_{53} + r_{\text{q}} = \reportednumber{\measureddfiveinelasticcollisionratevspressure}$ from a linear fit to the decay rate versus pressure.  This shifts the $\mathrm{D}_{5/2}$ state decay rate up by \reportednumber{\dfiveinelasticcollisionshifttotaldecayrate} at \systemnumber{\reportingdatapressure}.

For the $\mathrm{D}_{3/2}$ state, given our state detection scheme, any population transferred to the $\mathrm{D}_{5/2}$ state contributes to the dark state probability. Therefore, a separate measurement of $r_{35}$ is needed to calculate the $\mathrm{D}_{3/2}$ state decay rate shift due to fine structure mixing. We measure $r_{35} = \reportednumber{\threetofivemixingrate}$ at \systemnumber{\reportingdatapressure}, which shifts the $\mathrm{D}_{3/2}$ state decay rate down by \reportednumber{\threetofivemixingshiftindecayrate}~\cite{supp}. Instead of measuring the quenching rate with additional $\mathrm{D}_{3/2}$ lifetime measurements at elevated pressures, we bound the corresponding uncertainty with \reportednumber{\dfiveinelasticcollisionshiftdecayrateuncunit}, which is the combined $r_{53} + r_{\text{q}}$ uncertainty.

Systematic uncertainties with smaller effects are discussed in Supplement Material~\cite{supp}. Accounting for the shifts and uncertainties summarized in~\tref{table:decay_rate_systematics}, the $\mathrm{D}_{5/2}$ state decay rate is \reportednumber{\reportingdfiveovtwodecayrate}, giving a \reportednumber{\reportingdfiveovtwolifetime} lifetime, and the $\mathrm{D}_{3/2}$ state decay rate is \reportednumber{\reportingdthreeovtwodecayrate}, giving a \reportednumber{\reportingdthreeovtwolifetime} lifetime.

\begin{table}[h]
 \captionsetup{justification=myjust, singlelinecheck=false}
\caption{Shifts and uncertainties (in $10^{-3}\,\text{s}^{-1}$) for the $6\mathrm{d}\, ^{2}\mathrm{D}_{5/2}$ and $6\mathrm{d}\, ^{2}\mathrm{D}_{3/2}$ state decay rates.}
\label{table:decay_rate_systematics}
\begin{ruledtabular}
\begin{tabular}{lcc|cc}
    & \multicolumn{2}{c|}{$6\mathrm{d}\, ^{2}\mathrm{D}_{5/2}$} & \multicolumn{2}{c}{$6\mathrm{d}\, ^{2}\mathrm{D}_{3/2}$}  \\
    Source & Shift & Uncertainty & Shift & Uncertainty \\
    \midrule
    Statistical & $-$ & \reportednumber{\statisticaldfiveovtwodecayrateunc} & $-$ & \reportednumber{\statisticaldthreeovtwodecayrateunc} \\
    Elastic collisions & \reportednumber{\dfiveelasticcollisionshiftdecayrate} & \reportednumber{\dfiveelasticcollisionshiftdecayrateunc} & \reportednumber{\dthreeelasticcollisionshiftdecayrate} & \reportednumber{\dthreeelasticcollisionshiftdecayrateunc} \\
    Inelastic collisions & \reportednumber{$\dfiveinelasticcollisionshiftdecayrate$} & \reportednumber{\dfiveinelasticcollisionshiftdecayrateunc} & \reportednumber{\dthreeinelasticcollisionshiftdecayrate} & \reportednumber{\dthreeinelasticcollisionshiftdecayrateunc} \\
    Max. likelihood & $-$ & \reportednumber{\dfiveinputdecayrateanalysisuncertaintynounit} & $-$ & \reportednumber{\dthreeinputdecayrateanalysisuncertaintynounit} \\
    Thermal radiation & \reportednumber{$\blackbodydfivedecayrateshiftnounit$} & \reportednumber{\blackbodydfivedecayrateshiftuncnounit} & \reportednumber{\blackbodydthreedecayrateshiftnounit} & \reportednumber{\blackbodydthreedecayrateshiftuncnounit} \\
    \bottomrule \\[-7pt]
    \bf{Total} & \reportednumber{\totaldfiveshiftdecayrate} & \reportednumber{\totaldfiveshiftuncdecayrate} & \reportednumber{\totaldthreeshiftdecayrate} & \reportednumber{\totaldthreeshiftuncdecayrate}
\end{tabular}
\end{ruledtabular}
\end{table}

\begin{table*}[htbp]
\captionsetup{justification=myjust, singlelinecheck=false}
\caption{The removal energies of the low-lying states (in cm$^{-1}$) in
different approximations discussed in the text are presented. The theoretical total and experimental results are given in the rows $E_{\rm total}$ and $E_{\rm expt}$.
The difference between $E_{\rm total}$ and $E_{\rm expt}$ ~\cite{RalKraRea11} 
is presented (in percent) in the row labeled ``Difference (\%)'' $\Delta_X \equiv E_X - E_{\rm expt}$
}
\label{Tab:E}
\begin{ruledtabular}
\begin{tabular}{lrrrrr}
                                  &$7\mathrm{s}\,^2\mathrm{S}_{1/2}$&$6\mathrm{d}\,^2\mathrm{D}_{3/2}$&$6\mathrm{d}\,^2\mathrm{D}_{5/2}$&$6\mathrm{p}\,^2\mathrm{P}_{1/2}$&$6\mathrm{p}\,^2\mathrm{P}_{3/2}$ \\
\hline \\[-0.7pc]
$E_{\rm DHF}$                     &  $75\,898$          &   $62\,356$         &   $61\,593$         &    $56\,878$        &   $52\,906$    \\[0.2pc]
$E_{\rm LCCSD}$                   &  $82\,508$          &   $70\,186$         &   $68\,436$         &    $60\,865$        &   $55\,894$    \\[0.2pc]
$E_{\rm CCSDT}$                   &  $81\,894$          &   $69\,584$         &   $67\,926$         &    $60\,493$        &   $55\,597$    \\[0.2pc]
$\Delta E_{\rm Breit}$            &    $-19$          &      $62$         &      $87$         &      $-54$        &     $-13$    \\[0.2pc]
$\Delta E_{\rm QED}$              &    $-74$          &      $66$         &      $54$         &       $13$        &       $7$    \\[0.2pc]
$\Delta E_{\rm extrap}$           &     $37$          &     $127$         &     $115$         &       $26$        &      $22$    \\[0.4pc]
$E_{\rm total}$                   &  $81\,838$          &   $69\,839$         &   $68\,182$         &    $60\,478$        &   $55\,613$    \\[0.3pc]
$E_{\rm expt}$~\cite{RalKraRea11} &  $81\,843$          &   $69\,758$         &   $68\,100$         &    $60\,491$        &   $55\,634$    \\[0.1pc]
Difference (\%)                        &  $-0.00$          &   $0.12$          &    $0.12$         &    $-0.02$        &   $-0.04$    \\
\hline \\[-0.6pc]
${\Delta_{\rm DHF}}$      &  $-5945$          &   $-7402$         &   $-6507$         &    $-3613$        &   $-2728$    \\[0.2pc]
${\Delta_{\rm LCCSD}}$            &    $666$          &     $438$         &     $336$         &      $374$        &     $260$    \\[0.2pc]
${\Delta_{\rm CCSDT}}$            &     $51$          &    $-174$         &    $-174$         &        $2$        &     $-37$    \\[0.2pc]
${\Delta_{\rm total}}$            &     $-4$          &      $80$         &      $82$         &      $-13$        &     $-21$
\end{tabular}
\end{ruledtabular}

\end{table*}

{\it Theory}---We consider Ra$^+$ to be a monovalent ion and construct the finite basis set of one-particle orbitals in the $V^{N-1}$ approximation within the framework of the Dirac-Hartree-Fock (DHF) approach.
The Breit interaction and quantum electrodynamical (QED) corrections are also taken into account \cite{QED}.

We use the coupled-cluster single double triple (CCSDT) method, in which the coupled-cluster equations are solved iteratively, including the core and valence triple excitations~\cite{CCSDT}. In the equations for single and double cluster amplitudes, the sums in excited states were carried out with 45 basis orbitals with orbital quantum number $l \leq 6$.
In the equations for valence triples, we allowed the excitations of core electrons from the $[3\mathrm{s}-6\mathrm{p}]$ shells;
the sums in excited states were carried out with 32 basis orbitals with $l \leq 5$.

For an iterative solution of the equations for the core triples, more restrictions were applied due to drastically increased computational time. We solved these equations by allowing the core excitations from the $[4\mathrm{s}-6\mathrm{p}]$ core shells, the maximal orbital quantum number $l$ of all excited orbitals was equal to four, and the largest principal quantum number of the virtual orbitals where excitations were allowed was 22. 
These restrictions balance the enormous computational resources required for such a complete inclusion of the triple excitation with the need for high accuracy. We have verified that the present restrictions of these parameters give sufficient numerical accuracy by performing several computations with a different number of included core shells and virtual orbitals.

We started by calculating the removal energies of the low-lying states. The results of several increasing precision computations, as well as three additional corrections, are presented
in Table~\ref{Tab:E}. The lowest order DHF excitation energies are labeled ``DHF.'' We then perform the calculation within the framework of the linearized coupled-cluster single double (LCCSD) approximation. The most complete
calculation included the NL terms and the valence and core triple excitations. We designate this calculation as CCSDT.

We also list the QED corrections ($\Delta E_{\rm QED}$) and the corrections due to the Breit
interaction ($\Delta E_{\rm Breit}$) and basis extrapolation ($\Delta E_{\rm extrap}$). The latter is the contribution of the higher ($l > 6$) partial waves. It was determined based on previous studies \cite{Pr}. The total values, shown in the row labeled ``$E_{\rm total}$,'' are determined as the CCSDT values plus the three corrections. The difference between the total and experimental values
is given (in percent) in the row labeled ``Diff. (\%).''

To illustrate a consistent improvement in the results when we add different corrections,
we present the differences between the theoretical and experimental values obtained at each stage of the calculation in the lower panel of Table~\ref{Tab:E}.
Comparing $\Delta_{\rm total}$ with the experimental values ~\cite{RalKraRea11}, we see a very good agreement for the removal energies of the
$7\mathrm{s}\,^2\mathrm{S}_{1/2}$ and $6\mathrm{p}\,^2\mathrm{P}_{1/2,3/2}$ states. A slightly larger difference between theory and experiment for
the $6\mathrm{d}\,^2\mathrm{D}_{3/2,5/2}$ states is likely attributed to the nonlinear triple terms contribution omitted in our calculation as well as a larger contribution of the higher partial waves for these states. But even for $6\mathrm{d}\,^2\mathrm{D}_{3/2,5/2}$, the agreement with the experiment, at the level of 0.1\%, is exceptionally good for such a complicated system.

In~\tref{Tab:E2}, we present the reduced matrix elements (MEs) of the electric-quadrupole moment operator,
$\langle 7\mathrm{s}\,^2\mathrm{S}_{1/2} ||Q|| 6\mathrm{d}\,^2\mathrm{D}_{3/2,5/2} \rangle$, calculated in different approximations discussed above.
The results displayed in the rows labeled ``DHF'' and ``LCCSD'' are obtained in the DHF and LCCSD approximations, respectively.
Rows 3--6 give different corrections. Corrections resulting from NL terms and triples are listed in rows labeled ``$\Delta$(NL)'' and ``$\Delta$(Tr).''
The Breit interaction and QED corrections are small, and we present their total value in the row labeled ``$\Delta$(Breit \& QED)''.
To estimate the contribution of partial waves with $l > 6$, we reconstructed the basis set, including partial waves with the orbital quantum number up to $l= 7$. The difference between the LCCSD values of the MEs obtained for the basis sets with $l_{\rm max}= 7$ and $l_{\rm max}= 6$ is given in the row labeled ``$\Delta\,(l=7)$''. 
The final (recommended) values are obtained as the sum of the LCCSD values plus all corrections listed in rows 3--6.
We note that these corrections essentially cancel each other out and all of them have to be included in the precision computation. 

There are several sources of uncertainties in the final values of the MEs, such as small residual numerical inaccuracy in the calculation of the correlation corrections, omission of the NL terms in the triple equations, and a contribution from partial waves with $l>7$. Based on an estimate of possible contributions to the MEs from these effects, we assign uncertainties at the level of 0.5\% to the final values. 

\begin{table}[tp]
\captionsetup{justification=myjust, singlelinecheck=false}
\caption{Reduced MEs $\langle 7\mathrm{s}\,^2\mathrm{S}_{1/2} ||Q|| 6\mathrm{d}\,^2\mathrm{D}_{3/2,5/2} \rangle$ obtained in the DHF, LCCSD, and CCSDT approximations
(see text for details) are presented in $|e| a_0^2$, where $a_0$ is the Bohr radius. 
The uncertainties of the final values are
given in parentheses.}
\label{Tab:E2}%
\begin{ruledtabular}
\begin{tabular}{ldd}
                          &\multicolumn{1}{c}{$\langle ^2\mathrm{S}_{1/2} ||Q|| ^2\mathrm{D}_{3/2} \rangle$}
                                        & \multicolumn{1}{c}{$\langle ^2\mathrm{S}_{1/2} ||Q|| ^2\mathrm{D}_{5/2} \rangle$} \\
\hline      \\[-0.6pc]
    DHF                   & 17.26       & 21.77       \\[0.1pc]
   LCCSD                  & 14.59       & 18.69       \\[0.2pc]
$\Delta$(NL)              &  0.16       &  0.19       \\[0.1pc]
$\Delta$(Tr)              & -0.11       & -0.11       \\[0.1pc]
$\Delta$(Breit \& QED)    & -0.02       & -0.03       \\[0.1pc]
$\Delta\,(l=7)$           & -0.02       & -0.02       \\[0.1pc]
Final CCSDT                   & 14.60\,(7)  & 18.72\,(9)  \\[0.2pc]
Ref.~\cite{Pal2009}       & 14.74\,(15) & 18.86\,(17) \\[0.2pc]
Ref.~\cite{Sahoo2007}   & 14.87\,(7)  & 19.04\,(5)

\end{tabular}
\end{ruledtabular}
\end{table}

Using these values of the MEs, we calculated the electric-quadrupole and magnetic-dipole transition rates $W$, and lifetimes of the $^2\mathrm{D}_{3/2}$ and $^2\mathrm{D}_{5/2}$ states.
We note that $W(^2\mathrm{D}_{5/2} \rightarrow\, ^2\mathrm{D}_{3/2})$ is completely dominated by the $M1$ transition.
The contribution of the electric-quadrupole transition $^2\mathrm{D}_{5/2} \rightarrow\, ^2\mathrm{D}_{3/2}$  is very small,
and we neglect it. We find results (see~\tref{Tab:W}) that are in good agreement with those obtained in~\cite{Pal2009}.

\begin{table}[tp]
\captionsetup{justification=myjust, singlelinecheck=false}
\caption{Transition rates ($W$) of the electric-quadrupole $^2\mathrm{D}_{3/2,5/2} \rightarrow \, ^2\mathrm{S}_{1/2}$ and the
magnetic-dipole $^2\mathrm{D}_{5/2} \rightarrow\, ^2\mathrm{D}_{3/2}$ transitions and the lifetimes ($\tau$) of the
$^2\mathrm{D}_{3/2}$ and $^2\mathrm{D}_{5/2}$ states are presented. The uncertainties are given in parentheses.}
\label{Tab:W}%
\begin{ruledtabular}
\begin{tabular}{ccclcc}
&&&\multicolumn{1}{c}{This work} & \multicolumn{1}{c}{Ref.\cite{Pal2009}} & \multicolumn{1}{c}{Ref.\cite{Sahoo2007}} \\
\hline      \\[-0.6pc]
$W$(s$^{-1}$) & $E2$ & $^2\mathrm{D}_{3/2} \rightarrow\, ^2\mathrm{S}_{1/2}$  & $1.539(15)$ &  1.568     &           \\[0.3pc]
              & $E2$ & $^2\mathrm{D}_{5/2} \rightarrow\, ^2\mathrm{S}_{1/2}$  & $3.207(32)$ &  3.255     &           \\
              & $M1$ & $^2\mathrm{D}_{5/2} \rightarrow\, ^2\mathrm{D}_{3/2}$  &  0.049      &  0.049     &          \\[0.3pc]

$\tau$(ms)    &      & $^2\mathrm{D}_{3/2}$                            & $650(7)$    & $638(10)$  & $627(4)$   \\[0.1pc]
              &      & $^2\mathrm{D}_{5/2}$                            & $307(3)$    & $303(4)$   & $297(4)$
\end{tabular}
\end{ruledtabular}
\end{table}

Using the final values of the MEs given in \tref{Tab:E2} we find the ratio
\begin{equation}
R_{\rm E2} \equiv \left| \frac{\langle ^2\mathrm{S}_{1/2} ||Q|| ^2\mathrm{D}_{5/2} \rangle}
                              {\langle ^2\mathrm{S}_{1/2} ||Q|| ^2\mathrm{D}_{3/2} \rangle} \right| = 1.282\,(3) .
\label{RE2}
\end{equation}

We estimate the uncertainty of this ratio as the largest difference between the values of $R_{\rm E2}$
obtained in different approximations. Using Eq.~(\ref{RE2}), we find the ratio of the transition rates,
\begin{eqnarray}
R_{\rm W2} &\equiv& \frac{W(^2\mathrm{D}_{5/2} \rightarrow\, ^2\mathrm{S}_{1/2})}{W(^2\mathrm{D}_{3/2} \rightarrow\, ^2\mathrm{S}_{1/2})} \approx 2.084(6) .
\end{eqnarray}

Since the uncertainties of the transition energies are negligible compared to the uncertainty of $R_{\rm E2}$, we estimate
the absolute uncertainty of $R_{{\rm W}2}$ as $\Delta R_{{\rm W}2} \approx 2 (\Delta R_{{\rm E}2})= 0.006$.

\begin{figure}
    \centering
    \includegraphics[width=8.6cm]{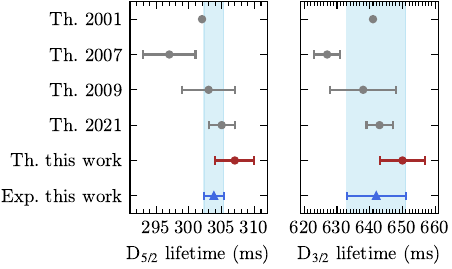}
    \captionsetup{justification=myjust, singlelinecheck=false}
    \caption{Comparison of the measured $\mathrm{D}_{5/2}$ state and $\mathrm{D}_{3/2}$ state lifetimes with theoretical calculations in this and previous works~\cite{Dzuba2001, Sahoo2007, Pal2009, Li2021a}.}
    \label{fig:theory_comp}
\end{figure}

A comparison of theoretical values and measured lifetimes is shown in Fig.~\ref{fig:theory_comp}.  No uncertainty is assigned to the reported lifetimes in~\cite{Dzuba2001}. We also note that the uncertainties reported in~\cite{Li2021a} appear to be underestimated. For example, Ref.~\cite{Li2021a} reported $\langle ^2\mathrm{S}_{1/2} ||Q|| ^2\mathrm{D}_{3/2} \rangle = 14.687(42)$ a.u. calculated in the framework of the CCSD method. Table~\ref{Tab:E2} of this Letter shows that the triple excitations correction for $\langle ^2\mathrm{S}_{1/2} ||Q|| ^2\mathrm{D}_{3/2} \rangle$, which is omitted in their computation, is $-0.11$, almost 3 times larger than the total uncertainty of 0.04 assigned in Ref.~\cite{Li2021a}.

{\it Conclusion}---We have measured and calculated the lifetimes of the $6\mathrm{d}\,^2\mathrm{D}_{3/2}$ and $6\mathrm{d}\,^2\mathrm{D}_{5/2}$ states of Ra$^+$.  This Letter presents the first tests of the CCSDT method for transition properties.  The long lifetimes of the $6\mathrm{d}\,^2\mathrm{D}_{5/2}$ and the $6\mathrm{d}\,^2\mathrm{D}_{3/2}$ states support the prospect of using $7\mathrm{s}\,^2\mathrm{S}_{1/2} \rightarrow 6\mathrm{d}\,^2\mathrm{D}_{5/2}$ and $7\mathrm{s}\,^2\mathrm{S}_{1/2} \rightarrow 6\mathrm{d}\,^2\mathrm{D}_{3/2}$ E2 clock transitions for future frequency standards with the Ra$^+$ ion.

{\it Acknowledgement}---The authors thank H. H\"{a}ffner for useful discussions. H.L. was supported by ONR Grant No.~N00014-21-1-2597 and M.F. was supported by DOE Award No.~DE-SC0022034. H.D., S.K., R.K., R.A.\,R., A.S., M.B., C.A.\,H. and A.M.\,J. were supported by the Heising-Simons Foundation Award No.~2022-4066, the W.M. Keck Foundation, NIST Award No.~60NANB21D185, NSF NRT Grant No.~2152201, the Eddleman Center, the Noyce Initiative, and NSF Grant No.~2326810, No.~2146555, and No.~1912665.  The isotope used in this research was supplied by the U.S. Department of Energy Isotope Program, managed by the Office of Isotope R\&D and Production.
The theoretical work was supported by the U.S. NSF Grant  No. PHY-2309254,  U.S. Office of Naval Research Grants No.~N00014-20-1-2513, No.~N000142512105 and the European Research Council (ERC) under the Horizon 2020 Research and Innovation Program of the European Union (Grant Agreement No. 856415).
The calculations in this work were done through the use of Information Technologies resources at the University of Delaware, specifically the high-performance Caviness and DARWIN computer clusters.  

{\it Data availability}---The data that support the findings of this Letter are openly available~\cite{Li2025}. The data supporting the theoretical results are available in the Letter.
 
\bibliography{references}

\end{document}


\title{Supplemental Material}
\maketitle

\section{Elastic Collisions}
Elastic collisions transfer kinetic energy to the trapped ion, which can Doppler broaden the cooling transition and remove the ion from the fiducial region. Due to these two effects, elastic collisions can decrease the scattered photon counts during state detection regardless of the ion's electronic state. For the $\mathrm{D}_{5/2}$ state lifetime measurement, this reduces the measured population in the ``bright'' ($\mathrm{S}_{1/2}$ and $\mathrm{D}_{3/2}$) states which systematically shifts up the $\mathrm{D}_{5/2}$ state lifetime. We model the population changes from both $\mathrm{D}_{5/2}$ state decays and elastic collisions, at rate $r_{\text{e}}$, using the following equations: 
\begin{align}
\label{eq:d5ov2-lifetime-elastic-collision-d5ov2}
    &\dot{P}_{5}(t) = -\frac{P_{5}(t)}{\tau_{5}} - r_{\text{e}}P_{5}(t) \\
\label{eq:d5ov2-lifetime-elastic-collision-s1ov2}
    &\dot{P}_{\text{b}}(t) = +\frac{P_{5}(t)}{\tau_{5}} - r_{\text{e}}P_{\text{b}}(t).
\end{align}
$P_{5}(t)$ is the population in the $\mathrm{D}_{5/2}$ state that did not experience an elastic collision. Therefore, it decreases with $\mathrm{D}_{5/2}$ state decays at a rate of 1/$\tau_5$, and $P_{5}(t)$ also decreases when the ion experiences an elastic collision (the $-r_{\text{e}}P_{5}(t)$ term). $P_{\text{b}}(t)$ is the population in the bright states that did not experience an elastic collision, which increases with $\mathrm{D}_{5/2}$ state decays and decreases due to elastic collisions.

For the $\mathrm{D}_{3/2}$ state lifetime measurement, population changes from $\mathrm{D}_{3/2}$ state decays and elastic collisions can be modeled with similar equations:
\begin{align}
\label{eq:d3ov2-lifetime-elastic-collision-d3ov2}
    &\dot{P}_{3}(t) = -\frac{P_{3}(t)}{\tau_{3}} - r_{\text{e}}P_{3}(t) \\
\label{eq:d3ov2-lifetime-elastic-collision-s1ov2}
    &\dot{P}_{\text{1}}(t) = +\frac{P_{3}(t)}{\tau_{3}} - r_{\text{e}}P_{\text{1}}(t).
\end{align}
$P_{3}(t)$ is the population in the $\mathrm{D}_{3/2}$ state that did not experience an elastic collision. Therefore, it decreases with $\mathrm{D}_{3/2}$ state decays at a rate $1/\tau_{3}$, and $P_{3}(t)$ also decreases when the ion experiences an elastic collision (the $-r_{\text{e}}P_3(t)$ term).  $P_{1}(t)$ is the population in the $\mathrm{S}_{1/2}$ state that did not experience an elastic collision, which increases with $\mathrm{D}_{3/2}$ state decays and decreases due to elastic collisions.

To account for systematic effects due to elastic collisions for the $\mathrm{D}_{5/2}$ lifetime, we model the measured dark state population as
\begin{equation}\label{eq:d5ov2-elastic-collision-fitting-model}
    P_{5 \text{d}}(t) = a(1-P_{\text{b}}(t)),
\end{equation}
where $a$ is the dark state population at zero delay.  Solving for $P_{\text{b}}(t)$ from Eqs.~(\ref{eq:d5ov2-lifetime-elastic-collision-d5ov2}) and (\ref{eq:d5ov2-lifetime-elastic-collision-s1ov2}) assuming the population is initialized in the $\mathrm{D}_{5/2}$ state, we get 
\begin{equation} \label{eq:d5ov2-lifetime-elastic-model-solution}
    P_{\text{b}}(t) =  e^{-r_{\text{e}} t} \times \bigl(1 - e^{-\frac{t}{\tau_{5}}}\bigr).
\end{equation}
Similarly, to account for systematic effects due to elastic collisions for the $\mathrm{D}_{3/2}$ state lifetime, we model the measured dark state population as
\begin{equation}\label{eq:d3ov2-elastic-collision-fitting-model}
    P_{3 \text{d}}(t) = a\bigl(1-P_{\text{1}}(t)-(1-\text{B}_{35})P_{\text{3}}(t)\bigr),
\end{equation}
where $\text{B}_{35}=\reportednumber{\branchratiodfiveovtwotosoneovtwo}\%$ is the branching fraction from the $\mathrm{P}_{3/2}$ state to the $\mathrm{D}_{5/2}$ state~\cite{Fan2019a} that accounts for optical pumping. Solving for $P_{\text{3}}(t)$ and $P_{\text{1}}(t)$ from Eqs.~(\ref{eq:d3ov2-lifetime-elastic-collision-d3ov2}) and (\ref{eq:d3ov2-lifetime-elastic-collision-s1ov2}) assuming the population is initialized in the $\mathrm{D}_{3/2}$ state, then
\begin{equation} \label{eq:d3ov2-lifetime-elastic-model-solution-1}
    P_{\text{1}}(t) = e^{-r_{\text{e}} t} \times \bigl(1 - e^{-\frac{t}{\tau_{3}}}\bigr)
\end{equation}
\begin{equation} \label{eq:d3ov2-lifetime-elastic-model-solution-2}
    P_{\text{3}}(t) = e^{-r_{\text{e}} t - \frac{t}{\tau_{3}}}
\end{equation}
We use Eqs.~(\ref{eq:d5ov2-elastic-collision-fitting-model}) and (\ref{eq:d3ov2-elastic-collision-fitting-model}) to fit for the $1/\tau_{5}$ and $1/\tau_{3}$ decay rates using the measured elastic collision rate $r_{\text{e}}$ at \reportednumber{\reportingdatapressure}. Their differences with the corresponding fitting results using the exponential decay function are calculated as the decay rate shifts due to elastic collisions.

\begin{figure}
    \centering
    \begin{subfigure}[b]{0.48\textwidth}
        \centering
        \includegraphics[width=\textwidth]{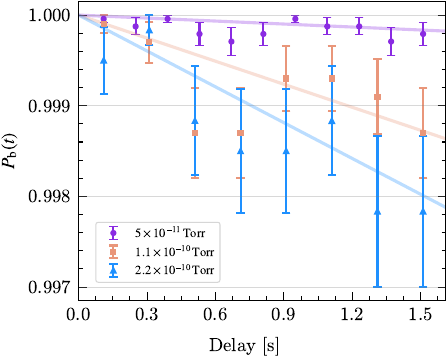}
        \caption{}
    \end{subfigure}
    \hfill
    \begin{subfigure}[b]{0.48\textwidth}
        \centering
        \includegraphics[width=\textwidth]{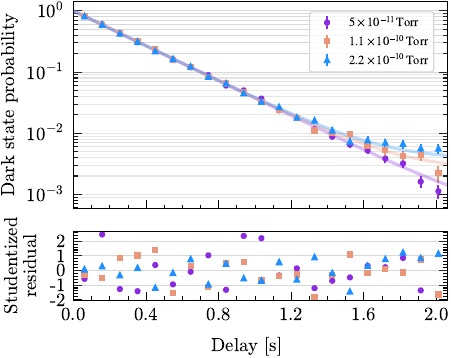}
        \caption{}
    \end{subfigure}
    \captionsetup{justification=myjust, singlelinecheck=false}
    \caption{(a) Elastic collision rate measurements taken at three different pressures,  \reportednumber{\reportingdatapressure} (purple circles), \reportednumber{\dfiveinelasticcollisionpressureone} (orange squares), \reportednumber{\dfiveinelasticcollisionpressuretwo} (blue triangles) which are fit to exponential decays.  (b) The measured dark state probability of the three $\mathrm{D}_{5/2}$ lifetime measurements are fit with Eq.~(\ref{eq:d5ov2-elastic-collision-fitting-model}) using the measured elastic collision rate, $r_{\text{e}}$, at the corresponding pressure.}
    \label{fig:d5ov2_lifetime_measurement_and_elastic_collision_rate_measurement_with_different_pressure}
\end{figure}

We measure the elastic collision rate by preparing the population in the $\mathrm{S}_{1/2}$ electronic ground state, waiting for a variable delay (\systemnumber{\elasticlowerdarktime} to \systemnumber{\elasticupperdarktime}), and performing state detection. When the population is initialized in the $\mathrm{S}_{1/2}$ state, the solution of Eq.~(\ref{eq:d5ov2-lifetime-elastic-collision-s1ov2}) is $P_{\text{b}}(t) = e^{-r_{\text{e}} t}$. The elastic collision rate is then extracted from an exponential fit of $P_{\text{b}}(t)$. The measurement is performed at three different background pressures and we obtain $r_{\text{e}}$ = \reportednumber{\elasticcollisionratereportingdatapressure} at \reportednumber{\reportingdatapressure}, $r_{\text{e}}$ = \reportednumber{\elasticcollisionrateelevatedpressureone} at \reportednumber{\dfiveinelasticcollisionpressureone}, and $r_{\text{e}}$ = \reportednumber{\elasticcollisionrateelevatedpressuretwo} at \reportednumber{\dfiveinelasticcollisionpressuretwo}, shown in Fig.~\ref{fig:d5ov2_lifetime_measurement_and_elastic_collision_rate_measurement_with_different_pressure}~(a).

\section{Inelastic Collisions}

\begin{figure}[h]
    \centering
    \includegraphics[width=8.6cm]{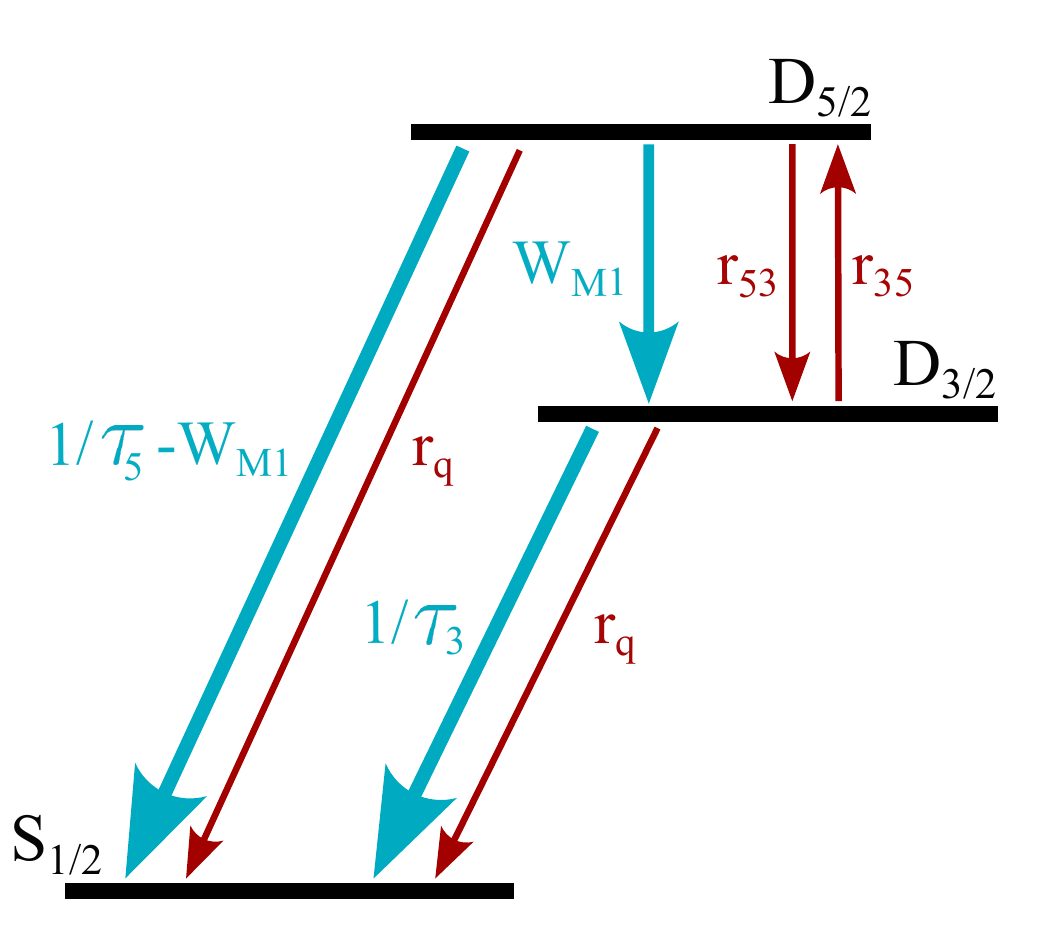}
    \captionsetup{justification=myjust, singlelinecheck=false}
    \caption{Radiative decays (blue arrows) and inelastic-collision induced transitions (red arrows) considered in the text.  The labels are the corresponding transition rates.}
    \label{fig:inelastic_collisions}
\end{figure}

Inelastic collisions between $^{224}$Ra$^+$ ions and background gas molecules may change the electronic state of the radium ions.  Two types of inelastic collisions may affect the lifetime measurements: 1) fine structure mixing collisions that transfer population between the $\mathrm{D}_{3/2}$ and $\mathrm{D}_{5/2}$ states, and 2) quenching collisions that drive population from the metastable D states to the electronic ground state.  We assume the quenching rate, $r_{\text{q}}$, is the same for the $\mathrm{D}_{5/2}$ and $\mathrm{D}_{3/2}$ states~\cite{Yu1997}, and we label the fine struture mixing rate from the $\mathrm{D}_{5/2}$ ($\mathrm{D}_{3/2}$) state to the $\mathrm{D}_{3/2}$ ($\mathrm{D}_{5/2}$) state as $r_{53}$ ($r_{35}$).  Fig. \ref{fig:inelastic_collisions} shows the spontaneous decays and collision-induced transitions considered here.

Inelastic collision effects have been studied for the D states of other alkaline earth ions including Ca$^+$~\cite{Knoop1995, Knoop1998} and Ba$^+$~\cite{Yu1997, KnabBernardini1992, Madej1990a}.  For Ra$^+$ the previous lifetime measurement of the $\mathrm{D}_{5/2}$ state reports $ r_{\text{q}} + r_{53} = \reportednumber{\versolatoradiumiondfiveinelasticcollisionratevspressure}$ with a background gas that is primarily neon~\cite{Versolato2011a}.  We assume that the background gas for our baked vacuum chamber is mostly hydrogen.  In measurements of other alkaline earth ions, the fine structure mixing and quenching rates are higher in inelastic collisions with hydrogen compared to neon, but the rates are not more than an order of magnitude faster~\cite{Knoop1998, Madej1990a}. There was no previous measurement of $r_{35}$ for Ra$^+$.  However, a fine structure mixing rate of the Ba$^+$ $\mathrm{D}_{3/2}$ state is reported (\systemnumber{\yubariumiondthreeinelasticcollisionrate}) in a lifetime measurement carried out at a similar background pressure~\cite{Yu1997}.  For Ra$^+$,  $r_{35}$ is expected to be smaller than the corresponding value for Ba$^+$ at the same background pressure.  This is because Ba$^+$'s fine structure splitting (\reportednumber{\bariumiondstateseparation}) is smaller than Ra$^+$'s (\reportednumber{\radiumiondstateseparation}) and the fine structure mixing rate decreases exponentially in the splitting~\cite{Knoop1995}.  Therefore, inelastic collision rates are expected to be small for Ra$^+$ at ultra-high vacuum, so we disregard the effects of multiple inelastic collisions.

\subsection{Inelastic collisions in the $\mathrm{D}_{5/2}$ state lifetime measurement}\label{subsec:d5ov2_inelastic_collisions}

\begin{figure}
    \centering
    \includegraphics[width=8.6cm]{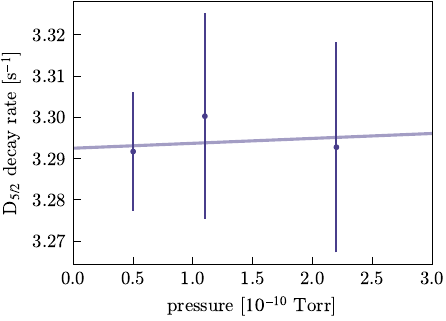}
    \captionsetup{justification=myjust, singlelinecheck=false}
    \caption{The linear fit of the $\mathrm{D}_{5/2}$ decay rate versus background pressure. From the fit we find the sum of fine structure mixing and quenching rates, $r_{53}+r_{\text{q}}$.  The fluctuation in the values is small compared to the statistical uncertainty, resulting in a reduced-chi square $\chi^2_\nu = \reportednumber{\dfiveinelasticcollisionratefitreducechisquare}$.}
    \label{fig:d5ov2_fit_inelastic_collision_rate}
\end{figure}

For the $D_{5/2}$ state lifetime measurement, the ion is initialized in the $D_{5/2}$ state. The decay rate from the $D_{5/2}$ state to the $S_{1/2}$ and $D_{3/2}$ states is $1/\tau_5 + r_{53} + r_\text{q}$. We measure the decay rate at two elevated pressures, \systemnumber{\dfiveinelasticcollisionpressureone} and \systemnumber{\dfiveinelasticcollisionpressuretwo}, see Fig. \ref{fig:d5ov2_lifetime_measurement_and_elastic_collision_rate_measurement_with_different_pressure}~(b). At higher pressures, the collisional quenching rate and the collisional fine-structure mixing rate will increase, while the spontaneous decay rate will remain constant. From a linear fit, we determine that the slope of the inelastic collision rate versus pressure is $\reportednumber{\measureddfiveinelasticcollisionratevspressure}$, see Fig.~\ref{fig:d5ov2_fit_inelastic_collision_rate}.  Inelastic collisions shift the $D_{5/2}$ state decay rate up by \reportednumber{\dfiveinelasticcollisionshifttotaldecayrate} at the base vacuum pressure of \reportednumber{\reportingdatapressure}.

\subsection{Inelastic collisions in the $\mathrm{D}_{3/2}$ state lifetime measurement}

For the $\mathrm{D}_{3/2}$ state lifetime measurement, the ion is initialized in the $\mathrm{D}_{3/2}$ state. For the collision-induced quenching rate, we note that $r_\text{q} < r_{53} + r_\text{q} = $ \reportednumber{\dfiveinelasticcollisionshifttotaldecayrate}.  Therefore, we bound $r_\text{q} $ to be within \reportednumber{\dfiveinelasticcollisionshiftdecayrateuncunit}. 

We determine the collision-induced fine structure mixing rate by initializing the ion in the $\mathrm{D}_{3/2}$ state and measuring the ion population after \systemnumber{\threetofivemixingratemeasurementdarktime} of delay. If a fine structure mixing collision occurs, the ion will be in the $\mathrm{D}_{5/2}$ state and appear dark.  Otherwise, the ion is in a bright state. This measures the sum of the fine structure mixing collision rate, $r_{35}$, and the elastic collision rate. We perform the same measurement with the ion initialized in the $\mathrm{S}_{1/2}$ state to measure the elastic collision rate. From the two measurements, we find $r_{35} = $ \reportednumber{\threetofivemixingrate} at \systemnumber{\reportingdatapressure}. 

We can model the population change of the D states as
\begin{align}
\label{eq:fine structure mixing-d5ov2}
    \dot p_{5}(t) &= -\frac{p_{5}(t)}{\tau_{5}} + r_{35} \cdot p_{3}(t) \\
\label{eq:fine structure mixing-d3ov2}
    \dot p_{3}(t) &= -\frac{p_{3}(t)}{\tau_{3}} - r_{35} \cdot p_{3}(t) - r_{\text{q}} \cdot p_{3}(t),
\end{align}
where $p_5$ ($p_3$) is the population in the $\mathrm{D}_{5/2}$ ($\mathrm{D}_{3/2}$) state. As the population in the $\mathrm{D}_{5/2}$ state must be small, further inelastic collisions after the $\mathrm{D}_{5/2}$ is populated are ignored. We evaluate the $\mathrm{D}_{3/2}$ state decay rate shift with $r_{35}$, and the statistical values for $\tau_3$ and $\tau_5$.  For the $\mathrm{D}_{3/2}$ lifetime measurement, our measured dark state probability can be expressed as $p_{5,\text{op}}(t) =\text{B}_{35} \cdot p_{3}(t) + p_{5}(t)$, where $p_{5,\text{op}}(t)$ denotes the $\mathrm{D}_{5/2}$ population after optical pumping.  We plug $p_3(t)$ and $p_5(t)$ solved from Eqs.~(\ref{eq:fine structure mixing-d5ov2}) and (\ref{eq:fine structure mixing-d3ov2}) into $p_{5,\text{op}}(t)$ to simulate the data.  The simulated data is fit to an exponential decay. The difference between the simulated data decay rate and the statistical decay rate is the shift due to fine structure mixing, \reportednumber{\threetofivemixingshiftindecayrate}.

\section{Analysis}
Maximum likelihood analysis requires an initial guess for the $\mathrm{D}_{5/2}$ state lifetime~\cite{Fan2019a}.  We use the theoretical lifetime $\tau_{5, \text{th}} = $ \reportednumber{\paldfiveovtwolifetime}~\cite{Pal2009}.  The uncertainty of $\tau_{5, \text{th}}$ results in an uncertainty of \reportednumber{\dfiveinputdecayrateanalysisuncertainty} for the $\mathrm{D}_{5/2}$ state decay rate and an uncertainty of \reportednumber{\dthreeinputdecayrateanalysisuncertainty} for the $\mathrm{D}_{3/2}$ state decay rate.

When fitting with an exponential decay, Eq.~(\ref{eq:d5ov2-elastic-collision-fitting-model}) and Eq.~(\ref{eq:d3ov2-elastic-collision-fitting-model}), the amplitude, $a$, is a fit parameter.  For the $\mathrm{D}_{3/2}$ lifetime measurement, this accounts for imperfect state preparation (SP2).  For the $\mathrm{D}_{5/2}$ lifetime measurement, although state detection two (SD2) is used as a filter, it shifts the delay time start.  This is because SD2 also filters out data that corresponds to decays during SD2, which means the delay time for the fitted data actually starts during SD2.  However, since the delay time shift is common to all data points it does not shift the fitted decay rate. By the same reasoning the delay time shift contributes to the uncertainty of $a$, the exponential fitting amplitude.

\section{Thermal Radiation}
The black-body radiation (BBR) stimulated emission rate for any electromagnetic
transition is
\begin{equation} \label{eq:blackbody-stimulated_emission-rate}
    r_{\text{bb}} = \frac{A_{21}}{e^{\nicefrac{\hbar\omega_{21}}{kT}} - 1},
\end{equation}
where $A_{21}$ is the spontaneous emission rate, $\omega_{21}$ is angular transition frequency, and $T$ is the black-body temperature~\cite{Siegman1986}. At \systemnumber{\roomtemperature}, the black-body radiation stimulated transition rates for optical transitions are negligible. For the Ra$^+$ $\mathrm{D}_{5/2} - \mathrm{D}_{3/2}$ M1 transition, given $A_{21}=\reportednumber{\dfivetodthreedecayrate} $~\cite{Pal2009}, $\omega_{21}=2\pi \times \reportednumber{\dfivetodthreetransitionfrequency}$~\cite{Kofford2025}, and $T = \reportednumber{\ionchambertemp}$, we calculate black-body radiation stimulated transition rate to be \reportednumber{\blackbodytransitionrate}. This can be directly subtracted to shift the $\mathrm{D}_{5/2}$ state decay rate. For the $\mathrm{D}_{3/2}$ state, with the calculated blackbody stimulated transition rate, the same analysis as the fine structure mixing effect is used to obtain a decay rate shift of \reportednumber{\blackbodydthreedecayrateshift}.

\section{Light Leakthrough}
The laser leakthrough from the acousto-optics modulator (AOM) double-passes is blocked by shutters (Stanford Research System, SR475) during the delay. The shutter is closed at the start of the delay time, and after the delay time, we wait an extra \systemnumber{\shutterwaittime} for the shutter to open before state detection. Because we expect the shutter to open within \systemnumber{\shutterwaittime}, there will be a short time that AOM laser leakthrough can interact with the ion.  Therefore, we measure the light leakthrough for all four lasers used in the experiment.  For example, to detect 1079~nm laser leakthrough, we prepare the ion in the $\mathrm{D}_{3/2}$ state, wait for a delay time of \systemnumber{\leakthroughdelaytime}, pump the remaining population to the $\mathrm{D}_{5/2}$ state and then apply state detection.  This measurement is done twice, once with the shutter open and again with the shutter closed, and we look for a difference in the dark state probabilities. We did not measure any change in the dark state probability.  Therefore, the associated systematic uncertainties are negligible compared to other reported uncertainties.

\section{Heating during delay time}
When a stray electric field displaces the ion from the rf null, the ion experiences excessive heating due to noise in the rf driving field, which can remove the ion from the fiducial region. To mitigate this effect, we carefully compensate for the stray field before each measurement run (typically \systemnumber{\compensationtime}). We also measure the effect of heating by leaving the ion in the dark for \reportednumber{\ionheatingdarktime} and then collecting fluorescence. We repeat this measurement more than \reportednumber{\ionheatingrepeattime} times and observe no delay in fluorescence with a \reportednumber{\ionheatingbintime} bin size of the collected photon counts. Therefore, the resulting systematic uncertainty from ion heating during the delay time can be neglected.

\section{Stray electric fields}
Uncompensated electric fields can push the ion off the rf null.  The rf field, which is effectively at dc compared to optical transitions, can mix opposite-parity states due to the dc Stark effect. If either D state mixes with the $\mathrm{P}_{1/2}$ or $\mathrm{P}_{3/2}$ states, the measured lifetimes will be reduced~\cite{Barton2000, Dijck2018}.  With the matrix elements for Ra$^+$~\cite{UDportal}, the dc field induced transition rate from the $\mathrm{D}_{5/2}$ state to the ground state is \reportednumber{\dfivedcstarkinducedtransitionrate} and the rate from the $\mathrm{D}_{3/2}$ state to the ground state is \reportednumber{\dthreedcstarkinducedtransitionrate}.  The typical field used for micromotion compensation is about \reportednumber{\compensationEfieldstrength}. The estimated residual stray field at the trap center is about \reportednumber{\strayEfieldstrength}.  These induced transition rates for the D states are negligible compared to other systematic uncertainties.

\section{Off-resonant scattering}
Both 468 nm and 1079 nm light can pump population out of the $\mathrm{D}_{5/2}$ state via off-resonant scattering.  The calculated combined deshelving rate is \reportednumber{\totaloffresonantscatteringrate}~\cite{Barton2000} with the laser intensities used during state detection.  Therefore, the shift of the measured dark state probability due to off-resonant scattering during the \systemnumber{\statedetecttime} state detection is negligible compared to other systematic effects.

\bibliography{references}